\documentclass[journal,onecolumn]{IEEEtran}
\usepackage[noadjust]{cite}
\newcommand\notype[1]{\unskip}

\ifCLASSINFOpdf
\usepackage[pdftex]{graphicx}
\graphicspath{{./Images/}}
\DeclareGraphicsExtensions{.pdf,.jpeg,.png}
\else

\usepackage[dvips]{graphicx}
\graphicspath{{./Images/}}
\DeclareGraphicsExtensions{.eps}

\fi
\usepackage{epstopdf}

\usepackage{amsmath}
\usepackage{amssymb}
\usepackage{bm}
\usepackage{bbold}
\usepackage{bbm}
\usepackage{mathrsfs}
\usepackage{algorithmic}
\usepackage{algorithm}

\usepackage{array}

\usepackage{enumitem}

\usepackage[caption=false,font=footnotesize]{subfig}

\usepackage{scrextend}
\usepackage{url}
\usepackage{xcolor}

\usepackage{multirow} %for columns spanning multiple rows in tables
\usepackage{booktabs,colortbl}

\usepackage{tabularx}
\newcolumntype{Y}{>{\raggedleft\arraybackslash}X}
\usepackage{siunitx}
\usepackage{blindtext}
\usepackage{verbatim} 
\usepackage{rotating}

\usepackage{float}
\floatstyle{ruled}
\newfloat{algorithm}{t}{lop}
\floatname{algorithm}{Algorithm}

\begin{document}
	\fontdimen2\font=0.6ex
	
\title{Strategies for Network-Safe Load Control with a Third-Party Aggregator and a Distribution Operator}

%Two Contrasting Strategies for Network-Safe Load Control that Use Operator-Aggregator Coordination}

%Two Strategies to Coordinate a Distribution Operator and a Third-Party Aggregator to Provide Network-Safe Load Control}


\author{Stephanie C. Ross
       and Johanna L. Mathieu% <-this % stops a space
\thanks{S. C. Ross and J. L. Mathieu are with the Department of Electrical Engineering
	and Computer Science, University of Michigan, Ann Arbor, MI 48109 USA
	(e-mail: sjcrock@umich.edu; jlmath@umich.edu).}% <-this % stops a space
\thanks{This work was supported by the Rackham Predoctoral Fellowship and by U.S. NSF Grant No. CNS-1837680.}%
}

%\markboth{Submitted to IEEE Transactions on Power Systems}%
%{Ross \MakeLowercase{\textit{et al.}}: Strategies for Network-Safe Load Control with a Third-Party Aggregator and a Distribution Operator}


\maketitle

\begin{abstract}
When providing bulk power system services, a third-party aggregator could inadvertently cause operational issues at the distribution level. We propose a coordination architecture in which an aggregator and distribution operator coordinate to avoid distribution network constraint violations, while preserving private information. The aggregator controls thermostatic loads to provide frequency regulation, while the distribution operator overrides the aggregator's control actions when necessary to ensure safe network operation. Using this architecture, we propose two control strategies, which differ in terms of measurement and communication requirements, as well as model complexity and scalability. The first uses an aggregate model and blocking controller, while the second uses individual load models and a mode-count controller. Both outperform a benchmark strategy in terms of tracking accuracy.  Furthermore, the second strategy performs better than the first, with only 0.10\% average RMS error (compared to 0.70\%). The second is also able to maintain safe operation of the distribution network while overriding less than 1\% of the aggregator's control actions (compared to approximately 15\% by the first strategy). However, the second strategy has significantly more measurement, communication, and computational requirements, and therefore would be more complex and expensive to implement than the first strategy.
\end{abstract}

%\begin{IEEEkeywords}
%Aggregation, demand response, distribution operation, frequency regulation,  thermostatically controlled loads
%\end{IEEEkeywords}

\section{Introduction}

High penetrations of aggregator-controlled distributed energy resources (DERs) pose a challenge to safe operation of distribution networks. In particular, DER aggregations participating in wholesale markets may cause unexpected power flows at the distribution level that result in reliability issues, as was noted by the U.S. Federal Energy Regulatory Commission (FERC) in a recent notice \cite{ferc_notice_2018}. This is of particular concern with third-party aggregators, which are separate from distribution operators and do not necessarily have awareness of the local impacts of their control actions. In the U.S., third-party load aggregators are active in ancillary service markets as a result of FERC Order No. 719 \cite{ferc_order_2008}. As an example of the problem, consider a third-party load aggregator participating in frequency regulation. To achieve a maximum setpoint, the aggregator may need 80\% of its loads to draw power at the same time---effectively turning a distribution network's diversified load into coincident load---which could cause violations of lower limits on voltage magnitudes~\cite{ross_effects_2018}. 

In this paper, we propose new strategies for controlling a particular type of load aggregation---an aggregation of residential, thermostatically controlled loads (TCLs). Examples of TCLs include air conditioners (ACs), heat pumps, and water heaters. The term ``TCL'' refers to both the thermal space (e.g., house) and appliance (e.g., (AC)). Every TCL has some thermal energy storage capacity, which allows for slight shifts in when electrical power is consumed without affecting temperature regulation. By controlling thousands of TCLs in this way, an aggregator can provide balancing services in the wholesale market~\cite{callaway_tapping_2009}. TCL aggregations are particularly well suited for services that are (or can be made) energy neutral, such as frequency regulation (i.e., secondary frequency control) \cite{pjm_neutrality_2017}. Most research on TCL control for fast balancing services has focused on the development of control strategies that 1) do not have a large impact on the TCLs' thermal regulation performance  and 2) have low measurement and communication requirements (e.g.,  \cite{mathieu_state_2013,zhang_aggregated_2013,kara_moving_2013,lu_evaluation_2012,hao_aggregate_2015,almassalkhi_packetized_2017, ruelens2017residential, radaideh2019sequential,liu2019trajectory}). Notably, all of the above-mentioned strategies have neglected distribution network operation and constraints.

%\rnew{Frequency regulation (or simply “regulation”) is the ancillary service provided by resources responding to the automatic generation control (AGC) signal. Frequency regulation is also referred to as secondary frequency control and load frequency control.}

	%Notably, all of the above-mentioned strategies neglect distribution network operation and constraints.}

%\rnew{Most research on TCL control for fast balancing services has neglected distribution network constraints, focusing instead on the development of control strategies that 1) do not perceptibly affect the TCLs' thermal regulation performance and 2) have low measurement and communication requirements. Many strategies \cite{callaway_achieving_2011,mathieu_state_2013,zhang_aggregated_2013,kara_moving_2013,lu_evaluation_2012,hao_aggregate_2015,almassalkhi_packetized_2017} ensure adequate thermal performance by switching TCLs on/off only when a TCL's temperature is inside its user-set range. In terms of measurement and communication requirements, \cite{mathieu_state_2013,zhang_aggregated_2013,kara_moving_2013} use aggregate modeling and estimation-based control techniques such that only an aggregate measurement is needed as feedback, whereas  \cite{lu_evaluation_2012,hao_aggregate_2015,almassalkhi_packetized_2017} require individual measurements or requests from each TCL. Notably, all of the above-mentioned strategies neglect distribution network operation and constraints.}

The objective of this paper is to develop TCL control strategies that provide frequency regulation without causing network constraint violations, all while preserving the privacy of the operator's and
third-party aggregator's private information. The distribution operator's network model and measurements are assumed private to ensure system security and consumer privacy, and the aggregator's control algorithm for providing frequency regulation is assumed private to maintain a competitive advantage commercially. Given these privacy constraints, we propose a coordination architecture in which the aggregator controls a TCL aggregation to provide frequency regulation, but, when necessary for network safety, the operator overrides the aggregator's commands to specific TCLs.

Our approach contrasts with that of prior work on network-safe load control, which has not focused on maintaining privacy and separation between the operator and aggregator. In \cite{vrettos_combined_2013}, the aggregator and operator are treated as the same entity, and a centralized AC optimal power flow (OPF) is solved. In \cite{dallanese_optimal_2018,hassan_optimal_2018,bernstein_real_2019}, distributed strategies are proposed, but the aggregator executes a prescribed (i.e., not private) algorithm that is coupled to the operator's algorithm through Lagrange multipliers. In \cite{molzahn_grid_2019}, a method for certifying if a distribution network will operate safely given any possible load control action is proposed; however, a control solution is still needed for networks that cannot be certified as safe, which are the types of networks we consider in this paper. Nazir and Almassalkhi are also investigating grid-aware approaches to dispatching demand \cite{nazir_convex_2019} and coordination of distributed energy resources \cite{nazir_inreview}; though the setting is not exactly the same as ours. 
 
%We develop and compare two control strategies with different attributes. Strategy I is easier to implement but lower accuracy than Strategy II. In Strategy I, the aggregator uses an aggregate-model based control strategy to provide \rnew{frequency} regulation, while the operator blocks particular TCLs from receiving the aggregator's commands when the network is at risk. In contrast, in Strategy II both the aggregator's and operator's control algorithms rely on individual models of each TCL; to ensure network safety, the operator not only blocks TCLs but also controls them with its own commands. In a case study, we evaluate the strategies in terms of regulation-signal tracking accuracy, and we test two versions of each control strategy. In the first, there is no coordination (i.e., no direct communication between operator and aggregator); in the second, the operator sends the aggregator partial information about its control actions. 

The main contributions of this paper are as follows. First, we propose a privacy-preserving coordination architecture that enables an aggregator and operator to coordinate to achieve network-safe TCL control. There are two variants of the architecture: with and without direct communication between operator and aggregator. Second, we propose two control strategies that use this coordination architecture. The strategies differ in terms of ease of implementation and accuracy, and thus provide options for operators and aggregators with different capabilities and preferences.
Third, we evaluate the performance of the proposed strategies in simulations of a realistic distribution network model with a high penetration of aggregated TCLs.

Certain aspects of the proposed control strategies relate to our prior work in~\cite{ross_coordination_2019,ross_managing_2019}. In \cite{ross_coordination_2019}, we proposed an alternative strategy for controlling a TCL aggregation to provide frequency regulation in a network-safe manner. This paper's Strategy I is similar to the strategy in \cite{ross_coordination_2019} in that they both use blocking to ensure network safety; the two strategies are different in all other respects, including coordination architecture as well as the aggregator's control algorithm. In \cite{ross_managing_2019}, we considered a different objective: providing local services to the distribution network by controlling small groups of TCLs. In this paper's Strategy~II, the operator uses the ``mode-count control algorithm'' from \cite{ross_managing_2019} to relieve voltage and line constraints. Lastly, \cite{ross_coordination_2019} and \cite{ross_managing_2019} did not include a full network model; in this paper, we test the strategies on a 613-node network model. 

%However, Strategy II's control architecture and the aggregator's control algorithm are new.

\section{Problem Description}

The {\em aggregator's objective} is to control the total power consumption of its TCL aggregation such that it tracks a frequency regulation signal with sufficient accuracy. We assume that the aggregator's control actions should be non-disruptive \cite{callaway_achieving_2011} to the TCL's end-user. Thus, we restrict an aggregator's control actions to switching TCLs on/off only when the TCLs are within their user-set temperature range like in \cite{mathieu_state_2013, zhang_aggregated_2013, kara_moving_2013, hao_aggregate_2015}. The aggregator's control is subject to the individual dynamics and constraints of each TCL.

A {\em distribution operator's objective} is to reliably deliver power of sufficient quality to consumers. In Table~\ref{tab:constraints}, we list the set of constraints for power quality and reliability that are considered in this paper. For transformers, apparent power is averaged over an hour because a transformer's thermal mass enables short-term violations of its power rating without causing overheating~\cite{mineraloil}. Note we will use the term ``network safety'' to refer to a distribution network's power quality and reliability.

Operators may need new tools to ensure network safety when third-party aggregators are active on their networks. Traditional network management tools, such as network reconfiguration and voltage regulation with regulators, have not been designed for the short but recurring fluctuations in power that can occur when aggregated loads provide frequency regulation. In this paper, we give the operator the ability to override the aggregator's control actions when necessary to avoid network constraint violations. As with the aggregator, we assume that the operator's TCL control actions must be non-disruptive to the end-user.   

%  Traditionally, distribution operators have had very few tools for real-time network management, relying instead on conservative planning based on peak load. 

The {\em control plant}, for both the aggregator and operator, is an aggregation of thousands of heterogeneous TCLs. Each TCL in the plant is modeled separately to capture its individual temperature dynamics and constraints. We use an individual TCL model that was developed in \cite{sonderegger_1978} and is commonly found in the literature (e.g.,\cite{mathieu_state_2013,hao_aggregate_2015}). For an aggregation of cooling TCLs, the $i$th TCL's temperature at time step $t{+}1$ is modeled~as
\begin{equation} \label{eq:tcl}
\begin{aligned}
\theta^{i}_{t+1} = 
&\begin{cases}
a^{i}\theta^i_t +\big(1-a^i\big) \big(\theta_{\textrm{a},t}-r^i p^i_{\theta}\big) & \hspace{-4pt} \text{if } m^i_t\hspace{-2pt}=\hspace{-2pt}1, \\  
a^i\theta^i_t+\big(1-a^i\big)\theta_{\textrm{a},t} & \hspace{-4pt} \text{if } m^i_t\hspace{-2pt}=\hspace{-2pt}0,
\end{cases}
\end{aligned}
\end{equation}
where $m$ represents the TCL's power status (1 for \emph{on} and 0 for \emph{off}) and $\theta_\text{a}$ is the ambient temperature outside of the TCL. Parameter $a^i$ is calculated as $a^i= \exp(-h/(c^i \, r^i))$ where $h$ is the duration of the discrete model's time step. All other parameters are defined in Table~\ref{tab:TCLparams}. The table also lists parameter values from \cite{mathieu_state_2013,mathieu_using_2012} for residential ACs. 

\begin{table} 
	\caption{Distribution Network Constraints}
	\centering
	\label{tab:constraints}
	\begin{tabular*}{.5\columnwidth}{ll rr}
		\toprule
		Component  &  Variable & Lower Limit & Upper Limit \\
		\midrule
		Line  & Current magnitude & -- & 100\% of rating \\
		Transformer & Avg. apparent power  & -- &  100\% of rating  \\
		Service node & Voltage magnitude  &0.95 p.u. &  1.05 p.u.   \\
		\bottomrule
	\end{tabular*}
\end{table}

\begin{table}
	\caption{Air Conditioner Parameters}
	\label{tab:TCLparams}
	\noindent
	\centering
	\begin{tabular*}{.38\columnwidth}{l r l}
		\toprule
		Parameter          & \centering Values     & Unit \\
		\midrule
		Thermal resistance ($r$) & 1.2-2.5       &   $^\circ$C/kW   \\
		Thermal capacitance ($c$) & 1.5-2.5 & kWh/$^\circ$C \\
		Thermal energy transfer rate ($p_{_\theta}$) & 10-18 & kW \\
		Setpoint temperature ($\theta_\text{set}$) & 18-27       &   $^\circ$C   \\
		Width of temperature range ($\delta$) & 0.25-1       &   $^\circ$C   \\
		Coefficient of performance ($\zeta$) & 2.5 & -- \\
		\bottomrule
	\end{tabular*}
\end{table}

To ensure non-disruptive control, we give the TCL's internal thermostat controller priority over external controllers. A TCL's power status is switched on/off by its internal controller if it reaches its upper or lower temperature limit. This condition is given by
\begin{equation} \label{eq:switching}
m^i_{t+1}=\begin{cases} 
1 & \text{if \ } \theta^i_t\geq\overline{\theta}^i,\\
0 & \text{if \ } \theta^i_t\leq\underline{\theta}^i, \\
\end{cases} 
\end{equation}
where $\underline{\theta}^i = \theta_\text{set}^i-\delta^i$ and $\overline{\theta}^i = \theta_\text{set}^i+\delta^i$ (see Table~\ref{tab:TCLparams} for parameter definitions).

TCLs with compressors (e.g., ACs) can also have a lockout constraint that prevents compressors from cycling too frequently. After a TCL's compressor is switched on/off, it is ``locked'' for a period of time before it can be switched again. Given a lockout period of $\tau_\text{L}$ time steps, a TCL's power status is constrained by
\begin{equation}\label{eq:lock}
m^i_t = m^i_{t-1} \ \ \text{if} \ \  \displaystyle\sum_{k = t-\tau_\text{L}}^{t-1}\mathbb{1}_{\{m^i_{t-1}\}}\big(m^i_k\big)  < \tau_\text{L} ,
\end{equation}
where the indicator function $\mathbb{1}_{\{f\}}\big(g)$ equals one if $g=f$ and zero otherwise. In this paper, we assume the lockout constraint is designed to prevent excessive switching by external controllers and that it does not apply to internal control. 

Finally, the total power of an aggregation of $N$ TCLs is given by $ P_{\text{total},t} = \sum_{i=1}^N p^i m^i_t$, where $p^i$ is a TCL's electric power rating with $p^i = p_\theta^i/\zeta^i$.

\section{Coordination Architecture}
\begin{figure}
	\centering
	\includegraphics[width=.6\columnwidth]{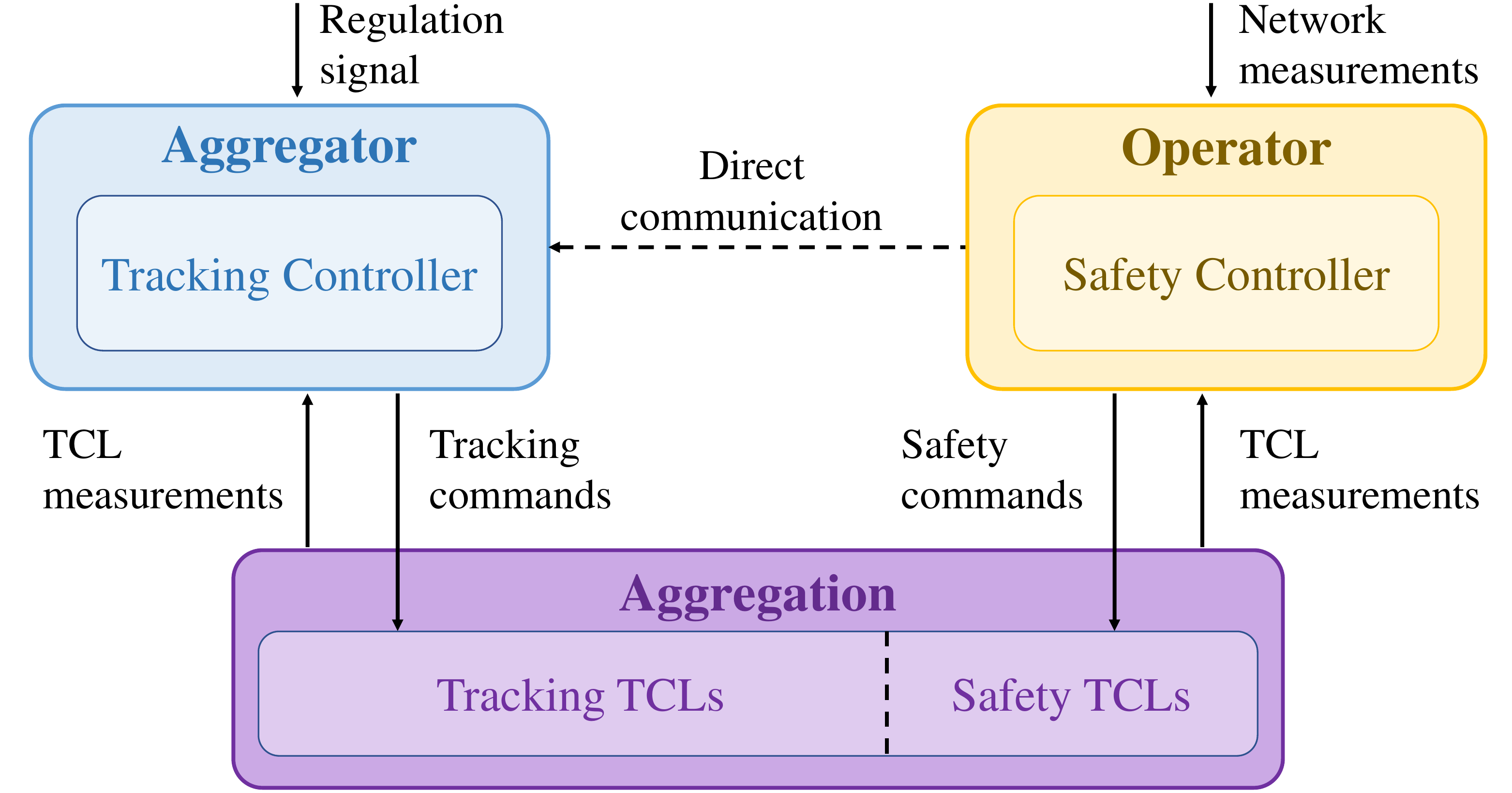}
	\caption{Coordination Architecture. The operator controls a portion of TCLs when necessary to prevent network violations. The aggregator controls all other TCLs to provide tracking. The dashed arrow indicates a variant in which the aggregator and operator directly communicate in real-time.}
	\label{fig:arch}
\end{figure}

We propose a coordination architecture by which the aggregator and operator can both achieve their control objectives while protecting their private information. Figure~\ref{fig:arch} shows the architecture: on the left side, the aggregator computes its next tracking command given TCL measurements and the regulation signal but does not have access to the operator's private information (i.e., network model and measurements); on the right side, the operator computes its next safety command based on network measurements (e.g., currents and voltages) and TCL measurements but does not have access to the aggregator's private information (i.e., its control algorithm and next control actions). It should be noted that optimal control could be achieved by sharing all information between the aggregator and operator and making centralized control decisions, but this would not achieve the privacy objective. 

%Because the operator's interventions occur in real-time, we assume the aggregator is unable to reduce its regulation capacity, remove TCLs from its aggregation, or adjust its contract with the wholesale market operator in any other way. This creates a challenge for the aggregator: the aggregator must provide the same \rnew{frequency} regulation service but with fewer TCLs responding to its control than expected.

We propose two control strategies---Strategy I and II---both of which use the proposed architecture. Strategy II is higher accuracy but harder to implement (with higher modeling, communication, and measurement requirements) than Strategy~I; Table~\ref{tab:strat} directly compares the strategies. In Strategy I, the aggregator uses an aggregate-model based control strategy to provide frequency regulation, while the operator blocks particular TCLs from receiving the aggregator's commands when the network is at risk. Since control partially synchronizes TCLs \cite{ross_effects_2018}, the aggregate power consumption of blocked TCLs is generally less variable than that of controlled TCLs. In Strategy~II both the aggregator's and operator's control algorithms rely on individual models of each TCL; in contrast to Strategy I, the operator \emph{actively} controls a subset of TCLs to reduce the variability of their aggregate power consumption to ensure network safety. Because Strategy~II relies on individual models of TCLs, it is less scalable: Strategy~II's models increase linearly with TCL-population size, while Strategy I's aggregate model is independent of population size. Strategy I is also more scalable in terms of communication requirements: the aggregator receives aggregated measurements and sends aggregated commands that are independent of population size. 

%Strategy~II generally has higher requirements than Strategy~I.

%Table~\ref{tab:strat} provides details on the components of each strategy and enables a comparison of their measurement and communication requirements. The aggregator's requirements are more substantial in Strategy II than Strategy I. In Strategy I, the aggregator requires only an aggregate TCL power measurement and sends back a small number of probabilistic commands to all TCLs. In contrast, in Strategy II, the aggregator requires measurements from each TCL and sends back individual on/off commands to each TCL. The operator's measurement and communication requirements are also more substantial in Strategy II than Strategy I. In Strategy I, the operator's controller requires only power measurements from each TCL and sends back a ``block'' or ``unblock'' command to each unit, as needed. In contrast, in Strategy II, the operator's controller requires temperature and power measurements from each TCL and sends back individual on/off commands to blocked units in addition to block/unblock commands. 

\begin{table}
	\renewcommand{\arraystretch}{1.1}
	\setlength{\tabcolsep}{5pt}
	\caption{Control Strategies (Agg. = Aggregator, Op. = Operator)}
	\label{tab:strat}
	\noindent
	\centering
	\begin{tabular*}{.5\columnwidth}{llllll}
		\toprule
		& & \hspace{-10pt} {Entity}  & {Controller} &   {Measurements} & {Commands to TCLs}   \\ 
		\midrule
		\multirow{4}{*}{\begin{sideways}Strategy I\end{sideways} } & \hspace{-10pt}\multirow{4}{*}{$\left\{\rule{0cm}{0.7cm}\right.$} & \hspace{-10pt} \multirow{2}{0.3in}{Agg.}   & \multirow{2}{0.79in}{Aggregate-Model Based Control} &  \multirow{2}{0.69in}{Aggregate power} & \multirow{2}{0.82in}{Probabilistic, one value for each bin} \\ \\ 
		
		& & \hspace{-10pt} \multirow{2}{0.3in}{Op.} & \multirow{2}{*}{Blocking Control} & \multirow{2}{0.69in}{Power \\ of each TCL} & \multirow{2}{0.82in}{Block or unblock to each TCL} \\ \\
		\midrule
		\multirow{4}{*}{\begin{sideways}Strategy II\end{sideways} }  &\hspace{-10pt}\multirow{4}{*}{$\left\{\rule{0cm}{0.7cm}\right.$} & \hspace{-10pt} \multirow{2}{*}{Agg.} & 
		\multirow{2}{0.8in}{Individual-Model Based Control}  & \multirow{2}{0.69in}{Power \& temp. of each TCL} & \multirow{2}{0.82in}{On/off to each TCL} \\ \\
		& & \hspace{-10pt} \multirow{2}{*}{Op.} & \multirow{2}{0.8in}{Mode-Count Control} & \multirow{2}{0.69in}{Power \& temp. of each TCL}  & \multirow{2}{0.82in}{On/off to each TCL}  \\ \\
		\bottomrule
	\end{tabular*}
\end{table}

%As we will see, block/unblock commands do not necessarily need to be sent every time step, whereas on/off commands to blocked units likely do -- thus communication will be more frequent in Strategy II than Strategy I.

\section{Methods: Strategy I}\label{sec:strategy1}

\subsection{Blocking Control}\label{sec:blocking}

In Strategy I, we propose that operators use blocking control to ensure safe network operation. When a TCL is blocked, it is unresponsive to aggregator commands, and it returns to its regular on/off cycles governed by its internal thermostat. To protect a particular network constraint, we block a group of TCLs whose coincident demand would otherwise contribute to the violation of the constraint. Blocking a group of TCLs reduces their on/off synchronization---and the resulting peaks and valleys in coincident demand---that would otherwise be caused by the aggregator's commands. 

To protect a network in real-time, we must identify the set of constraints that are at risk due to aggregator-control and the appropriate TCLs to block to reduce this risk. This problem is non-trivial: it needs to be solvable in real-time while accounting for 3-phase unbalanced power flow and network constraints, TCLs' discrete operating states (on and off), uncertainty in the aggregator's future control actions, and uncertainty in the uncontrolled load consumption at each bus. To the best of our knowledge, solution methods for this type of problem do not yet exist in the literature. Work on a related problem \cite{molzahn_grid_2019}
shows promise but relies on simplifying assumptions (e.g.,
balanced power flow enabling consideration of a single-phase
equivalent network). Our preliminary work on this problem~\cite{ross_method_2020} develops an OPF-type approach to constrain the norm of the vector of nodal power deviations caused by an aggregator, but it is conservative (in a deterministic setting) and does not consider uncertainty. Ref.~\cite{nazir_convex_2019} poses the idea of constraining dispatchable demand as a hosting capacity problem. It uses convex inner approximations of network-feasible power flows, but again assumes a deterministic setting.

Developing an online solution to this problem is important future work, but a significant undertaking on a different type of problem than the one considered here, and so considered outside the scope of this paper. Instead, we determine the set of TCLs to block ensuring network safety for the time horizon of the problem (e.g., an hour) using iterative offline three-phase unbalanced power flow simulations in which perfect knowledge of the full problem, including the aggregator's control actions over the time horizon, is assumed. Details of the method are given in Appendix~\ref{app:block}. Therefore, our results provide an upper-bound on performance; in practice, the distribution operator would not be able to select exactly the right set of TCLs to block (maximizing TCL control capacity while ensuring satisfaction of all network constraints) because of uncertainty.
%\rnew{Developing an online method is future work. To the best of our knowledge, solution methods for this type of problem do not yet exist in the literature, and work on a related problem~\cite{molzahn_grid_2019} shows promise but relies on simplifying assumptions (e.g., balanced power flow enabling consideration of a single-phase equivalent network).}

%Note we make the simplifying assumption that line-distance is proportional to the losses induced by a load; we also neglect TCLs' heterogeneous power ratings.

\subsection{Aggregate-Model Based Tracking Control} \label{sec:proposedtracking}

Strategy I's proposed tracking controller is based on an aggregate model of TCL state progression. In Section~\ref{sec:casestudy}, we benchmark the tracking performance of the proposed, aggregate-model based controller against that of a model-less, PI controller. We expect the proposed controller to have better performance because it uses model-based prediction to compensate for the internal control actions of TCLs and uses state estimation to prioritize switching TCLs that are about to internally switch.

\subsubsection{Aggregate TCL Model} 
The proposed model takes the form of a population transition model, which has been commonly used in the literature on TCL control (e.g., \cite{mathieu_state_2013,zhang_aggregated_2013,liu_model_2016, radaideh2019sequential,ji2019data}). In this type of model, the state space of all TCLs is discretized into the same number of ``bins'' and the progression of TCLs from bin to bin is modeled with transition probabilities.

\begin{figure}
 	\centering
 	\includegraphics[width=.6\columnwidth]{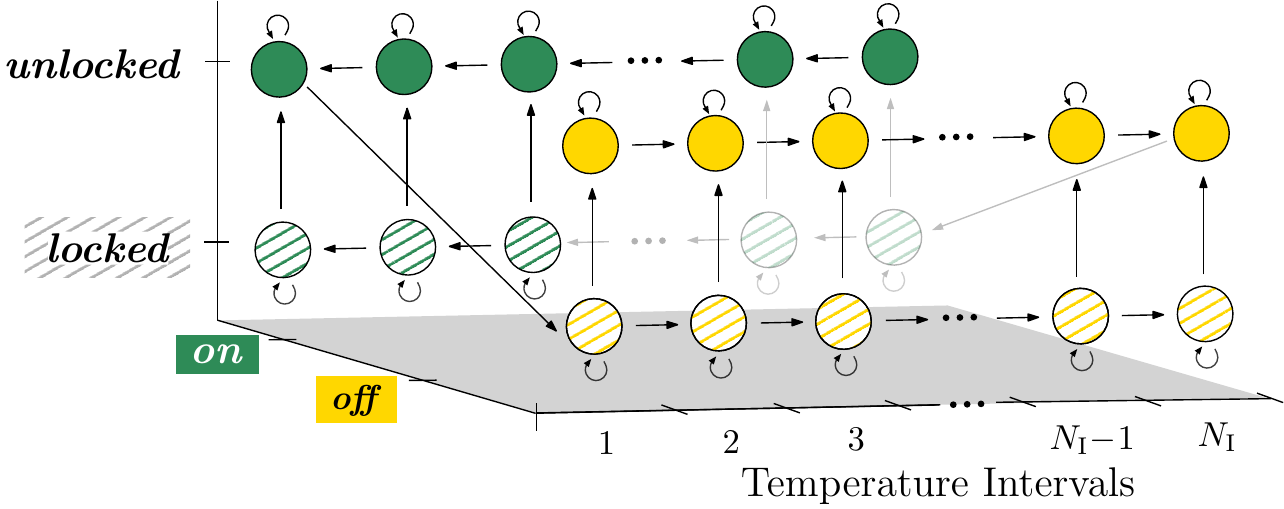}
 	\caption{State Transition Diagram with Locked States. Arrows indicate the most likely state transitions for an uncontrolled TCL with a lockout period.}
 	\label{fig:markov}
 \end{figure}

In this paper, we extend the population transition model to include TCL lockout dynamics. Figure~\ref{fig:markov} depicts the proposed model. Each bin is represented by a circle and is defined by a locked/unlocked status, on/off status, and one of $N_\text{I}$ temperature intervals. In Fig.~\ref{fig:markov}, horizontal arrows represent transitions from one temperature interval to the next due to TCLs' temperature dynamics. Diagonal arrows represent transitions due to  thermostat actions (i.e., internal control). Vertical arrows represent TCLs becoming unlocked after their lockout times have elapsed. Transitions due to aggregator-control are not shown but would be downward diagonals in each temperature interval.

We formulate the aggregate model as a linear time-varying system: 
 \begin{equation} \label{eq:ltv_kf}
\begin{aligned}
\bm{x}_{t+1}&=\bm{A}_{t}\bm{x}_{t}\\
\bm{y}_t&=\bm{C}\bm{x}_t,
\end{aligned}
\end{equation}
where $\bm{x}$ is the state vector, $\bm{A}$ the state transition matrix, $\bm{y}$ the output, and $\bm{C}$ the output matrix. The state $\bm{x}$ is the distribution of the TCL population across the bins. Given the different types of bins, the state vector can be divided into four categories: $\bm{x}=
[\underbrace{x^1  \ldots  x^{N_\text{I}}}_\text{off, unlocked} \ \ \underbrace{x^{N_\text{I}+1} \ldots x^{2N_\text{I}}}_\text{on, unlocked} \underbrace{x^{2N_\text{I}+1} \ldots x^{3N_\text{I}}}_\text{off, locked} \ \ \underbrace{x^{3N_\text{I}+1} \ldots x^{4N_\text{I}}}_\text{on, locked}]^T\vspace{2pt}$. The entries of the state transition matrix $\bm{A}_t$ are the probability that a TCL will transition from bin to bin. For example, the entry in the $i$th row and $j$th column is the probability that a TCL in bin $j$ will transition to bin $i$ in the next time step. 

We model transitions due to the aggregator's probabilistic commands within the $\bm{A}$ matrix. This is similar to the Markov Decision Process approach in \cite{chen_state_2017}. We formulate $\bm{A}_t$ as the product of two transition matrices:
$\bm{A}_t = \bm{A}_{\text{u,}t}\bm{A}_\text{s}$. Matrix $\bm{A}_{\text{u,}t}$ models ``external transitions'' that occur due to the TCLs' response to the aggregator's command $\bm{u}_t$. Matrix $\bm{A}_\text{s}$ models ``internal transitions'' that occur due to the TCLs' autonomous dynamics, including temperature dynamics, lockout dynamics, and internal control. For the purpose of modeling, we have assumed that, in a given time step, internal transitions occur before external transitions. The matrix $\bm{A}_\text{s}$ can be identified for a particular outdoor temperature by counting bin transitions when aggregator-control is inactive. Here, for simplicity, we assume a constant outdoor temperature, which makes $\boldsymbol{A}_\text{s}$ time-invariant and is not an unreasonable assumption over the duration of a frequency regulation market period, e.g., 1 hour.

Transition probabilities in $\bm{A}_{\text{u},t}$ depend on the aggregator's command $\bm{u}_t$. We design the aggregator's command to be a vector that can be broadcast to all TCLs and thereby reduce communication requirements. The $i$th entry of $\bm{u}_t$ is the probability with which TCLs in bin $i$ should switch at time step~$t$. The entries only correspond to bins that are unlocked, so $\bm{u}_t$ has length $2N_\text{I}$. As an example, consider a probabilistic command in which $u_5 = 0.4$, i.e., “TCLs in bin 5 switch with 40\% probability”. Each TCL receives this command, checks its current state, and, if in bin 5, decides whether to switch by running a Bernoulli random trial with 40\% chance of success (i.e., of switching). Given this definition of $\bm{u}_t$, the matrix $\bm{A}_{\text{u},t}$ takes the form
\begin{equation*}
\left[
\begin{array}{cc|cc} 
\hspace{-3.5pt}\bm{I}-\text{diag}(u^1_t, \dots, u_t^{N_\text{I}})  & \bm{0} & \bm{0}  & \bm{0} \\
\bm{0}  & \hspace{-9pt}\bm{I}-\text{diag}(u^{N_\text{I}+1}_t \hspace{-4pt}, \dots, u_t^{2N_\text{I}})  &  \bm{0} & \bm{0} \\
\bm{0}  & \hspace{-8pt} \text{adiag}(u^{2N_\text{I}}_t, \dots, u_t^{N_\text{I}+1}) &  \bm{I} & \bm{0} \\
\text{adiag}(u_t^{N_\text{I}}, \dots, u^1_t)  & \bm{0}&  \bm{0} & \bm{I}
\end{array}  
\right], 
\end{equation*} 
% $\bm{U}^{1}_t = \bm{I}-\text{diag}(u^1_t, \dots, u^{N_\text{I}})$, \mbox{$\bm{U}^{2}_{t} = \bm{I}-\text{diag}(u^{N_\text{I}+1}_t \hspace{-4pt}, \dots, u^{2N_\text{I}})$}, $\bm{U}^{3}_{t} = \text{adiag}(u^{2N_\text{I}}_t, \dots, u^{N_\text{I}+1})$, and $\bm{U}^{4}_{t} = \text{adiag}(u^{N_\text{I}}, \dots, u^1_t)$. 
where $\bm{I}$ is the identity matrix, diag$(\cdot)$ maps the input vector to a diagonal matrix, and adiag$(\cdot)$ maps the input vector to an anti-diagonal matrix, where the diagonal runs from the upper right corner to the lower left corner. The left side of $\bm{A}_{u,t}$ models transitions due to aggregator control: the diagonal sub-matrices model transitions out of unlocked bins, and the anti-diagonal sub-matrices model transitions into locked bins.

The two outputs of the aggregate model are: 1) the total power consumption of the aggregation and 2) the fraction of TCLs in the aggregation, which should always equal one. The output matrix is given by $\bm{C} = \left[ \bm{C}_\text{s} \ \ \bm{C}_\text{s} \right] $, where 
\begin{equation*}
\bm{C}_\text{s} = 
\left[
\begin{array}{cccccc}
0 & \ldots & 0 & \overline{p}_\text{on}N & \ldots & \overline{p}_\text{on}N   \\
1 & \ldots & 1 & 1 & \ldots & 1 \\
\end{array}  
\right]
\end{equation*}
and $\overline{p}_\text{on}$ is the average power consumption of an \emph{on} TCL.

\subsubsection{State Estimation}
We use a time-varying Kalman filter \cite{matlab_kalman_2018} to estimate the aggregate state in \eqref{eq:ltv_kf}. The output of the Kalman filter in each time step is $\hat{\bm{x}}_t$. We account for process noise and measurement noise by adding terms $\bm{w}$ and $\bm{v}$ to the state and output equations in  \eqref{eq:ltv_kf}, respectively. We treat the covariances of $\bm{w}$ and $\bm{v}$, denoted as $\bm{Q}$ and $\bm{R}$, as tuning parameters. For output measurements, we use $\bm{y}_{\text{meas},t} =  
\left[
\begin{array}{cc}
P_{\text{total},t} & 1  \\
\end{array}  
\right]^\text{T}.$
We assume the second output measurement is perfect (i.e., has zero measurement noise) and set the corresponding entry in $\bm{R}$ to zero. In this way, the second output equation acts as an equality constraint within the Kalman filter \cite{simon_kalman_2010}. 

\subsubsection{Control Policy} \label{sec:ctrlpolicy}
We propose a control policy based on one-step model prediction. The policy selects a control input such that the model's predicted output in the next time step matches the desired output. Appendix~\ref{ap:derivation} provides a derivation of the policy. We note that the proposed policy is similar to that of \cite{mathieu_state_2013}; slight differences emerge due to differences between our aggregate model and that of \cite{mathieu_state_2013}.

Here we describe the policy for when an increase in power is needed. First, the choice of $\bm{u}_t$ must satisfy the equation:
\begin{equation}
 \sum_{n=1}^{N_\text{I}} u_t^n x_{\text{pred},t}^{n}  = K \frac{P^*_{\text{total},{t+1}}-\bm{C}^1\bm{A}_\text{s}\hat{\bm{x}}_t}{\overline{p}_\text{on}N},
\label{eq:control_on1}
\end{equation}
where $\bm{x}_{\text{pred},t}=\bm{A}_\text{s}\hat{\bm{x}}_t$, $K$ is a proportional gain, $P^*_\text{total}$ is the desired power in the next time step, and $\bm{C}^1$ is the first row of $\bm{C}$. Because \eqref{eq:control_on1} is under-determined when $N_\text{I}>1$, we have the freedom to choose $\bm{u}$ as long as \eqref{eq:control_on1} is satisfied. We design a rule for selecting $\bm{u}$ that prioritizes switching TCLs that are closest to being switched by internal control. When switching TCLs \emph{on}, bin $N_\text{I}$ has first priority, bin $N_\text{I}{-}1$ has second priority, and so on. Let $b$ be the minimum index for which $\sum^{b}_{k=0} x_{\text{pred,}t}^{N_\text{I}-k}$ is greater than the right hand side of \eqref{eq:control_on1}. Then we set the entries of $\bm{u}$ equal to 1 for all bins of higher priority than $N_\text{I}{-}b$ and equal to zero for all bins of lower priority. Finally, the value for bin $N_\text{I}{-}b$ is chosen such that \eqref{eq:control_on1} is satisfied. 

The control policy is very similar when a decrease in power is needed. In this case, $\bm{u}$ must satisfy \eqref{eq:control_off} in Appendix~\ref{ap:derivation}, and bins $2N_\text{I}$ to  $N_\text{I}{+}1$ are switched out of, in that order.

%\vspace{-5pt}
%\subsection{Coordination}

Finally, when there is direct communication in Strategy I, the operator informs the aggregator of the number of TCLs it is blocking in the current time step. To incorporate this information into the aggregator's control policy, we use a lookup table of proportional gains for each of five blocking levels: 0\%, 10\% 20\%, 30\%, and 40\%. When the number blocked changes, we update $K$ in \eqref{eq:control_on1} by linearly interpolating between the nearest values in the lookup table. 

\section{Methods: Strategy II} \label{sec:strategy2}

\subsection{Mode-Count Control} \label{sec:modecount}
For the operator in Strategy II, we propose the use of mode-count control to ensure network safety. Mode-count control was originally developed for TCLs in \cite{nilsson_class_2017} and was extended to account for TCLs with lockout in \cite{ross_managing_2019}. In this paper, we propose two applications of mode-count control: 1) prevention of under-voltages or over-currents by reducing a TCL group's maximum demand, and 2) prevention of over-voltages by increasing a TCL group's minimum demand. For brevity, we present only the first application; the methods for the second application follow directly. 

For a group of co-located TCLs, the policy determines which TCLs to switch \emph{on}/\emph{off} to constrain the group's ``on-count'', and thereby the group's demand. A group's on-count is the number of TCLs in the group that are \emph{on} at a particular time and is denoted  $H^j_t$ for group $j$ at time $t$. We constrain a group's on-count such that $H^j_t\leq\overline{H}^j$,
where $\overline{H}^j$ is a low (but feasible) upper bound.

\begin{figure}
	\centering
	\includegraphics[width=.5\columnwidth]{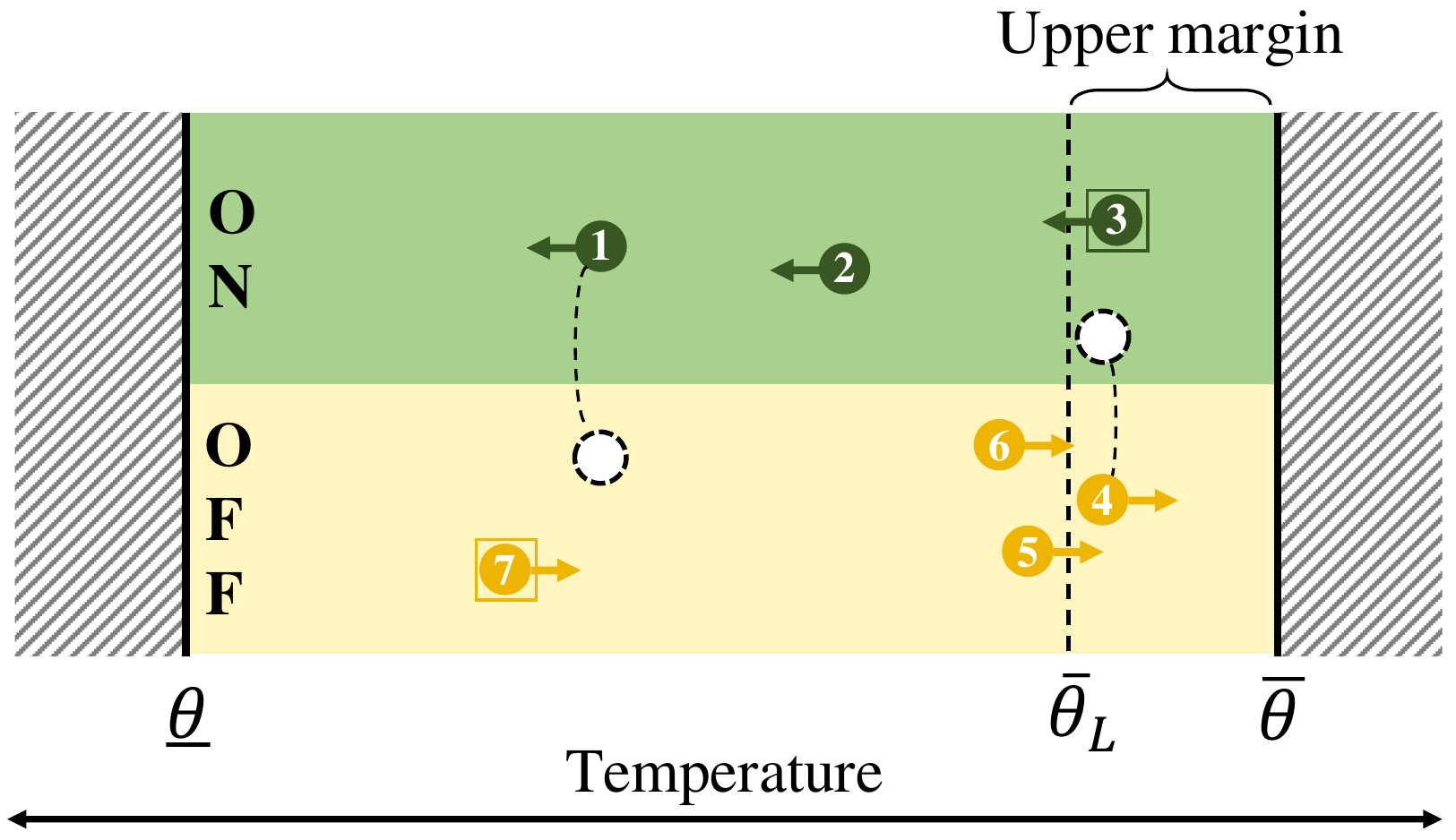}
	\caption{Demonstration of Mode-Count Control. The policy maintains the on-count $\leq 3$ by switching \emph{off} TCLs \emph{on}, as soon as possible after entering the upper margin, and if needed, switching an \emph{on} TCL \emph{off} to compensate. TCLs are indicated by numbered circles; boxes indicate locked status.}
	\label{fig:policy}
\end{figure}

Figure~\ref{fig:policy} provides an example of the control policy. Consider a group of 7 TCLs with upper bound $\overline{H}=3$. The central idea of the policy is to switch TCLs \emph{on} as soon as possible once they have entered the ``upper margin'' of their temperature range. At the time step shown in Fig.~\ref{fig:policy}, the policy switches TCL \#4 \emph{on} because it has just entered its upper margin, and switches TCL~\#1 \emph{off} to satisfy the counting constraint. A main goal of the policy is to avoid having too many TCLs locked \emph{on} at the same time, which could happen if we waited to switch TCLs at their upper temperature limit. 
%Consider what would happen if instead the policy waited for \emph{off} TCLs to reach $\overline{\theta}$ before switching them. Then, with the group shown in Fig.~\ref{fig:policy}, TCL \#'s 3,4, and 5 may all be locked \emph{on}, when \#6 switches \emph{on}, which would violate the upper bound $\overline{H}=3$. 

We summarize the steps of the control policy in Algorithm~1 and define three terms used within the algorithm as follows. (For a more detailed treatment of the policy, we refer the reader to~\cite{ross_managing_2019}.) First, a TCL's ``time-to-upper-limit'' is the time it would take a TCL to progress from its current temperature $\theta^i_t$ to its upper limit $\overline{\theta}^i$ in the \emph{off} direction. This time can be explicitly calculated from the individual TCL model, and for the $i$th TCL is given by $t^i_\text{UL} = r^ic^i \ln\big((\theta_{\text{a}}-\theta^i_t)/(\theta_{\text{a}}-\overline{\theta}^i)\big)$. Second, a TCL's ``upper margin temperature'' $\overline{\theta}_\text{L}^{i}$ is the lesser of two temperatures: 1) the temperature from which it takes the TCL $\tau_\text{L}$ time steps to reach $\overline{\theta}^{i}$ when \emph{off}, and 2) the temperature reached by the TCL $\tau_\text{L}$ time steps after leaving $\overline{\theta}^{i}$ when \emph{on}. Third, a TCL ``counter-switches'' with another TCL by switching at the same time but in the opposite direction (i.e., to the opposite power status). 

\begin{algorithm}
\small
In each time step, 
\begin{enumerate}[leftmargin=0.45cm]
	\item Initialize $\Delta{H} = \overline{H}-H_t$, and find the \emph{off} unit in the upper margin with shortest time-to-upper-limit. Let this unit's index be $g$.
	\item If unit $g$ is \emph{unlocked}: 
	\begin{enumerate}[leftmargin=0.45cm]
		\item If $\Delta{H}>0$, then switch unit $g$ \emph{on} and decrement $\Delta{H}$.
		\item Otherwise, find a TCL that is available to counter-switch; this TCL must be \emph{on} and \emph{unlocked}, must have $\theta^i_t<\overline{\theta}_\text{L}^i$, and must not be reserved to counter-switch with a different TCL. Let $s$ be the index of the available TCL with the longest time-to-upper-limit. Switch unit $g$ \emph{on} and unit $s$ \emph{off}.
	\end{enumerate}	 
	\item If unit $g$ is \emph{locked}:
	\begin{enumerate}[leftmargin=0.42cm]
		\item If possible, find a TCL that is available to counter-switch. Reserve this TCL for counter-switching with $g$ in the future.
		\item Otherwise if $\Delta{H}>0$, decrement $\Delta{H}$ so that unit $g$ will be able to switch in the future, as soon as it is unlocked.
	\end{enumerate}
	\item Find the \emph{off} unit with the next shortest time-to-upper limit. Let this unit's index be $g$.
	\item Repeat 2)-4) until either there are no more \emph{off}, \emph{unlocked} units in the upper margin, or $\Delta H = 0$ and there are no more \emph{on} TCLs available for counter-switching.
\end{enumerate}
\caption{Mode-Count Control for Upper Bound}
\end{algorithm}

To prevent the violation of a particular network constraint, an operator must first identify the group of TCLs to control and then set a lower bound on the group's on-count. As with blocking, we use an iterative offline method to identify the groups of TCLs to control for network safety. Once identified, we set each group's upper bound $\overline{H}^j$ to its lowest feasible value; our prior work \cite{ross_managing_2019} provides a conjecture on this value.

\subsection{Individual-Model Based Tracking Control}\label{sec:indmod}

For the aggregator in Strategy II, we propose a control policy based on individual TCL models. The general principal of the control policy is to switch TCLs that are closest---in terms of time---to being switched internally. The proposed controller is similar to the priority stack method of \cite{hao_aggregate_2015}, but we use time to determine priority instead of temperature. We summarize the steps of the control policy in Algorithm 2 and discuss each step in the remainder of the section. 

In step 1, we calculate the change in power that is needed to achieve tracking in the next time step. When there is no direct communication between aggregator and operator, this change in power is given by
\begin{equation}\label{eq:ptrack}
\Delta P_{\text{track},t} = P^*_{\text{total},t+1}-(P_{\text{total},t} + \Delta P_\text{internal});
\end{equation}
see the end of this section for the case in which there is direct communication. Variable $\Delta P_\text{internal}$ is the change in power that will occur in the next time step due to internal switching. We determine $\Delta P_\text{internal}$ by predicting which TCLs will be switched by their internal controllers in the next time step. For each TCL, we calculate its time-to-upper-limit $t_\text{UL}$ (as defined in Section~\ref{sec:modecount}) and its time-to-lower-limit $t_\text{LL}$, defined as the time it would take a TCL to progress from its current temperature to its lower limit in the \emph{on} direction. Both metrics can be solved for using \eqref{eq:tcl}. A TCL is predicted to switch in the next time step if it is \emph{off} and has $t_\text{UL}<h$, or if it is \emph{on} and has $t_\text{LL}<h$. 

\begin{algorithm}
	\small
	In each time step,
	\begin{enumerate}[leftmargin=0.45cm]
		\item Calculate $\Delta P_{\text{track},t}$, the change in power needed to track $P^*_{\text{total},t+1}$, taking into account changes in power due to internal switching and, if there is coordination, safety control.
		\item Update priority stacks $\mathbb{S}_{\text{on},t}$ and $\mathbb{S}_{\text{off},t}$.
		\item If $\Delta P_{\text{track},t}\geq P_\text{small}$
		\begin{enumerate}[leftmargin=0.45cm]
			\item Find index $j^*\in \mathbb{S}_{\text{off},t}$ that minimizes \mbox{$|\sum_{i=1}^j p^i-\Delta P_{\text{track},t}|$}
			\item Switch \emph{on} unit $j^*$ and all units of higher priority in $\mathbb{S}_{\text{off},t}$
		\end{enumerate}
		\item Otherwise, if $\Delta P_{\text{track},t}\leq -P_\text{small}$
		\begin{enumerate}[leftmargin=0.45cm]
			\item Find index $j^*\in \mathbb{S}_{\text{on},t}$ that minimizes $|\sum_{i=1}^j p^i+\Delta P_{\text{track},t}|$
			\item Switch \emph{off} unit $j^*$ and all units of higher priority in $\mathbb{S}_{\text{on},t}$
		\end{enumerate}	
	\end{enumerate}
\caption{Individual-Model Based Tracking Control}
\end{algorithm}

In step 2, we update the priority stacks. Priority stack $\mathbb{S}_\text{on}$ is composed of \emph{on} and \emph{unlocked} TCLs and is sorted by time-to-lower-limit; priority stack $\mathbb{S}_\text{off}$ is composed of \emph{off} and \emph{unlocked} TCLs and is sorted by time-to-upper-limit. In each time step, we update the priority stacks according to the switching actions from the last time step and newly calculated values for $t_\text{UL}$ and $t_\text{LL}$. The set of TCLs under control by the operator are excluded from the priority stacks; we assume the aggregator is able to identify this set given its TCL measurements.

%(Note, for TCLs that were not switched in the last time step, $t_\text{UL}$ and $t_\text{LL}$ can be updated by just subtracting the duration of a time step from the relevant metric.) 

In step 3, we select which TCLs to switch. To prevent excessive switching, we only switch TCLs if the desired change in power has magnitude greater than a threshold, $P_\text{small}$. Parameter $P_\text{small}$ can be thought of as a tuning parameter; here we set it to 25\% of the smallest, individual power rating in the aggregation. If the threshold is reached, then in step 3a we calculate how much of the stack should switch to minimize the tracking error. In this calculation, we take the sum of the TCLs' power ratings in order of priority. In step 3b, the selected TCLs are switched.

Step 4 is similar to step 3, except $\Delta P_\text{track}$ is negative, so units need to be switched \emph{off} rather than \emph{on}.

%\subsection{Coordination and Implementation}
Finally, when there is direct communication in Strategy II, the operator sends the aggregator $\Delta P_\text{safety}$, the change in power its safety control actions will cause in the next time step. The aggregator compensates for the operator's actions by subtracting $\Delta P_\text{safety}$ from the right-hand side of \eqref{eq:ptrack}. All other aspects of the control are the same.

%\rnew{One of the challenges of implementing Strategy II is that both the aggregator's and operator's control algorithms rely on having an accurate thermal model of each individual TCL. These individual models can be estimated using parameter identification techniques. If we assume each TCL's setpoint, temperature range, and coefficient of performance are known, then only parameters $r^i$ and $c^i$ must be identified. A nonlinear least-squares algorithm that uses temperature and on/off state histories can be used for this purpose \cite{ledva_managing_2017}.}
%The proposed controller is similar to the priority stack method of \cite{hao_aggregate_2015}. The largest difference between the methods is our use of time to determine priority instead of temperature. While a time-based priority stack always switches the TCL that would next switch naturally, this is not guaranteed with a temperature-based priority stack. Thus, a time-based priority stack reduces the total number of switching events, which is important in terms of user acceptance, as well as for control accuracy (more switching events results in more locked, uncontrollable TCLs). The downside to using time is that it requires an individual model of each TCL. However, such a model can easily be identified from a TCL's temperature and power measurements. Lastly, having these models enables the prediction of the change in power due to natural switching, $\Delta P_\text{natural}$, which improves tracking accuracy. 

\section{Case Study}\label{sec:casestudy}

\subsection{Benchmark Strategy} \label{sec:benchmark}

In the case study, we compare the proposed control strategies to a benchmark strategy. The benchmark is identical to Strategy~I, except the aggregator's tracking controller is a model-free proportional-integral (PI) controller. Therefore, comparison to the benchmark allows us to determine the benefit of the model-based controller and state estimator in Strategy I. The PI controller computes a scalar switching command, whose magnitude is the probability with which TCLs should switch and whose sign is the direction with which TCLs should switch: positive (\emph{on}) or negative (\emph{off}). We use a discrete-time PI control algorithm with anti-windup (see pp. 311-312 of \cite{astrom_feedback_2012}). The tuning parameters are a proportional gain $K_\text{P}$ and an integral gain $K_\text{P}/T_\text{I}$, where $T_\text{I}$ is the integral time constant. We set the anti-windup time constant equal to that of the integral gain. When there is direct communication, we use a look-up table for the gains.
%\begin{equation}
%\begin{aligned}
%\tilde{u}_{\text{\tiny PI},t} &= K_\text{P}(P^*_{\text{total},t}-P_{\text{total},t}) + e_t  \\
%u_{\text{\tiny PI},t} &= \sat(\tilde{u}_{\text{\tiny PI},t}) \\
%\end{aligned}
%\end{equation} 
%where $K_\text{P}$ is the proportional gain, $e_t$ is the integral term, and $\sat(\cdot)$ is a saturation function that limits the magnitude of the argument to 1. The integral term is updated in each time step according to
%\begin{equation}
%e_{t+1} = e_t+\frac{K_\text{P}}{T_\text{I}}(P^*_{\text{total},t}-P_{\text{total},t}) + \frac{1}{T_\text{T}}(u_{\text{\tiny PI},t}-\tilde{u}_{\text{\tiny PI},t}),
%\end{equation}
%where $T_\text{I}$ scales the integral gain relative to the proportional gain and $T_\text{T}$ scales the anti-windup term.

\subsection{Setup}    
We test the control strategies in simulations of a distribution network with a high-penetration of aggregator-controlled residential ACs. We assume 100\% of houses on the network have ACs, and we use an outdoor temperature of 32$^\circ$C in order to capture a peak-load hour. We compare the strategies' performances across 10 \mbox{1-hour} trials, where each trial uses a different segment of the RegD regulation signal from the PJM Interconnection \cite{regd}. We scale the signal such that its amplitude is +/-33\% of the ACs' baseline power consumption. Good control performance is indicated by high accuracy in tracking and low prevalence of network violations.

The simulation model has two main parts: 1) the network and 2) the AC aggregation and controllers. We use a realistic network model and GridLAB-D \cite{gridlabd} to solve the network's three-phase, unbalanced power flow. The network model is of an actual system: R1-12-47-1 in Pacific Northwest National Lab's (PNNL's) database \cite{pnnl_github,pnnl_taxonomy}. The network model includes dynamic voltage regulation in the form of one voltage regulator and two capacitor banks. During simulations, we measure current flow through lines, apparent power flow through transformers, and voltages at residential meters. Appendix~\ref{ap:feedermod} lists a few modifications made to the network model.

We model the AC aggregation and controllers in MATLAB. We determine the number of houses served by each residential transformer using a disaggregation method developed by PNNL~\cite{pop_script}. Each house is composed of one AC model and one constant-power baseload. We use \eqref{eq:tcl}-\eqref{eq:lock} to model each AC and generate a heterogeneous population by randomly selecting parameter values from the uniform distributions described in Table~\ref{tab:TCLparams}. The lockout period for each AC is set to 2 minutes. A power factor of 0.97 is used for ACs and 0.95 for baseload. Table~\ref{tab:aggregations} reports the AC aggregation's baseline power consumption, as well as other network and load metrics. 

\begin{table} 
	\caption{Case Study: Aggregation and Network Details}
	\label{tab:aggregations}
	\centering
	\begin{tabular*}{.5\columnwidth}{cccc}
		\toprule
		\# of Primary Nodes & \# of Houses & Baseline AC Load & Non-AC Load \\
		\midrule
		613 & 2265 & 4.42 MW & 2.01 MW  \\
		\bottomrule
	\end{tabular*}
\end{table}

For Strategy I's aggregate load model, we use five temperature intervals, i.e., $N_\text{I}=5$, and identify the matrix $\bm{A}_\text{s}$ by counting bin transitions within historical data of TCLs' temperatures and on/off statuses. For Strategy II's individual TCL models, we assume perfect models (i.e., the models are identical to those in the simulated plant); in future work, we plan to identify the individual TCL models from historical measurements.
 
We tune the tracking controllers of Strategy I and the benchmark strategy for the best average tracking performance over a set of tuning trials. When there is no direct communication, we use 50 tuning trials consisting of 25 different hours of the RegD signal and two different blocking levels: 0\% and 20\%. Table~\ref{tab:tuning} lists the resulting tuned parameters. When there is direct communication, we tune the controllers for each blocking level independently. The tuned parameters have the following ranges of values: $K \in[1,1.15]$, $K_\text{P}\in[3\times10^4, 5\times10^5]$, and $T_\text{I} \in [80, 800]$, where the lower and upper values correspond to 0\% and 40\% blocked, respectively. 

%, again using 25 hours of the RegD signal
%\rnew{For each strategy and trial, we determine which ACs should be under safety control by iterating through different sets of AC in preliminary simulations until no violations occur.}
%For each strategy and trial, we determine which ACs should be under safety control through a series of preliminary simulations. Before each simulation, we select additional ACs for safety control according to the methods proposed in Section~\ref{sec:blocking}. We repeat this process until no violations occur.

%In all cases, the trials used for tuning are different from those used for testing
\begin{table}
	\caption{Tuning Parameters}
	\label{tab:tuning}
	\noindent
	\centering
	\begin{tabular*}{.5\columnwidth}{llll}
		\toprule
		Strategy I: &    $K = 1 $ & $\bm{Q} = \text{diag}(I, 10^2I)$ &  $\bm{R} = 10^9 \text{diag}(1, \ 0)$ \\ 
		 Benchmark: &   $T_\text{I} = 100$ & $K_\text{P} = 3.4\times10^{-4}$ \\
		\bottomrule
	\end{tabular*}
	\\	diag$(\cdot)$ maps the input matrices to a block diagonal matrix
\end{table}

\subsection{Results}

Table~\ref{tab:results4} shows the results of the case study, with values averaged over the 10 test trials. For all strategies, the safety controllers are successful---\emph{no network violations occur}. Without safety control, there are an average of 2.2 over-voltage nodes and 1.2 overloaded lines, where the averages are taken over all trials and strategies. The number of ACs under safety control varies considerably, depending on the strategy used. In the benchmark strategy, 13.8\% of ACs are blocked, on average, in order to avoid network violations; this number falls slightly to 10.3\% when there is direct communication. Similarly, in Strategy I, 15.3\% of ACs are blocked, and 15.2\% with direct communication. In contrast, Strategy II's mode-count controller requires only 0.4\% of the aggregation to prevent network violations, with and without direct communication.

\begin{table}
	\renewcommand{\arraystretch}{1.1}
	\caption{Case Study Results: Averages over 10 Trials}
	\label{tab:results4}
	\noindent
	\centering
	\begin{tabular*}{.5\columnwidth}{l r r r r r}
		\toprule
		\multirow{2}{1cm}{Strategy}   &  \multicolumn{2}{c}{ACs under Safety Control} & & \multicolumn{2}{c}{Tracking Error (RMS)}  \\
		\cmidrule(){2-3} \cmidrule(){5-6}
		   & No Comm. & Comm. & & No Comm. & Comm.   \\
		\midrule
		Benchmark  &  13.8\%  & 10.3\%  &  &   1.37\% & 1.28\% \\
		Strategy I   &   15.3\% & 15.2\%  & &   0.72\% & 0.70\% \\
		Strategy II & 0.4\% &  0.4\%   & & 0.10\% & 0.10\% \\
		\bottomrule
	\end{tabular*}
\\ ``Comm.'' indicates direct communication between aggregator and operator
\end{table}

In terms of tracking error, Strategy II outperforms Strategy I, and both outperform the benchmark strategy. In Table~\ref{tab:results4}, root-mean-square (RMS) error in tracking is reported as a percentage of the baseline power of the aggregation.   Comparing against the benchmark, we find that the tracking error of Strategy I (0.72\%, and 0.70\% with direct communication) is almost 50\% less than that of the benchmark (1.37\%, and 1.28\% with direct communication), and the tracking error of Strategy II (0.10\% with and without direct communication) is more than an order of magnitude less than that of the benchmark. In all strategies, tracking performance improves given direct communication; we note that for Strategy II this improvement is not shown by Table~\ref{tab:results4} due to rounding. To put these numbers in context, we can compare them to those reported in the literature. For example, \cite{mathieu_state_2013} reported RMS errors of 0.2-9\% depending upon the communication scenario and  \cite{olama2018coordination} reported RMS errors of 0.5-28\% depending upon the number of TCLs participating, though neither study considered distribution network constraints.

Although Strategy II outperforms Strategy I in terms of tracking, Strategy I's advantage is in ease of implementation. Strategy~I is more computationally scalable than Strategy II because Strategy I's aggregate model is independent of aggregation size whereas Strategy II uses individual TCL models and therefore the model increases linearly with aggregation size. In the case study, we expect the benchmark strategy to be the fastest since it is model-free, Strategy I to be the next fastest with its 20-state model, and Strategy II to be the slowest with its 4530-state model. All strategies were run in MATLAB using a 4-core machine (3.30 GHz) with 32 GB of RAM. Over the 10 trials without coordination, the average time to compute commands for the next time step was $0.045\times10^{-4}$~s for the benchmark strategy, $1.5\times10^{-4}$~s for Strategy I, and $7.9\times10^{-4}$~s for Strategy II. All of the strategies are very fast given that the frequency regulation signal changes every two seconds. We expect Strategy I's computational advantages over Strategy II to be more pronounced with larger sized aggregations. Recall that Strategy I also requires substantially lower measurement and communication requirements than Strategy II (see Table~\ref{tab:strat}).

\begin{figure}
	\centering
	\includegraphics[width=.6\columnwidth]{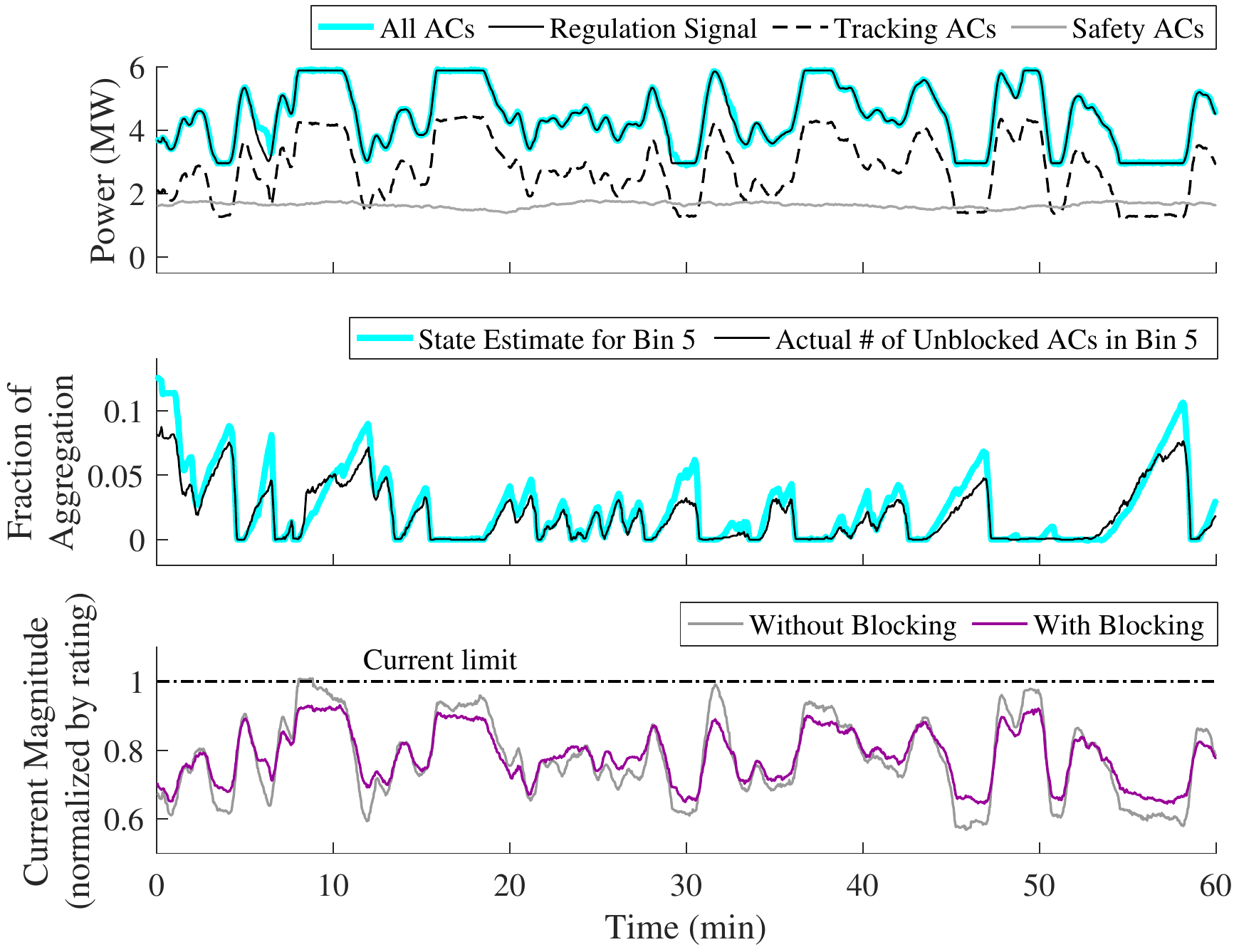}
	\caption{Strategy I during a selected trial. The plots shows the strategy's high tracking accuracy (top plot) and state-estimation performance for bin 5 (middle plot). The bottom plot shows that blocking a large portion of TCLs can prevent over-currents, here on overhead line \# 301.}
	\label{fig:3plot_ROM}
\end{figure}

In Fig.~\ref{fig:3plot_ROM}, we demonstrate a few of Strategy~I's key features using data from a selected trial. In the upper plot, Strategy~I's aggregate-model based controller closely tracks the regulation signal, despite a large percentage---in this trial, 38\%---of ACs being blocked. However, the tracking is not perfect: there is noticeable error in the $6^\text{th}$ minute, when the aggregation has saturated (i.e., not enough ACs are available to switch \emph{off} due to blocking or lockout).  The middle plot of Fig.~\ref{fig:3plot_ROM} demonstrates an attribute of the state estimator: for bins that frequently receive external switching commands (i.e., bins $N_\text{I}$ and $2N_\text{I}$), the state estimate corresponds to the number of ACs that are available to switch (i.e., unblocked ACs), rather than the total number of ACs in the bin. This helps the controller compensate for blocking. The lower plot of Fig.~\ref{fig:3plot_ROM} demonstrates the effects of blocking. Without blocking, phase C of overhead line 301 experiences an over-current in minutes 7 and 8; blocking a portion of the ACs on phase~C effectively reduces this over-current. 

%\rnew{This trial's RMS tracking error is 1.93\%, which is higher than the \mbox{10-trial} average of 0.72\% but still well within the performance standards of regional transmission organizations in the U.S. (e.g., see \cite{pjm_manual11_2019})}.

Fig.~\ref{fig:3plot_ROM} also highlights the main drawback of using blocking for safety control: a relatively large percentage of ACs must be blocked to protect just a few network constraints. This is because, unlike mode-count control, blocking does not actively constrain a group's maximum demand. Instead, blocking returns a group to its normal variations in demand and level of load diversity, where there is more load diversity with larger groups. %; this approach is more effective for larger groups of ACs with more diversified demand.    %Because larger groups of ACs have more diversified load, blocking is more effective with larger group sizes.

% Blocking works on the principle of load diversification: while tracking control partially synchronizes ACs, blocking enables ACs to return to their normal level of diversity.
%\rnew{Blocking works for two reasons: 1) with a large enough group, the loads' diversity results in a lower maximum demand than when the group is partially synchronized by tracking control, and 2) the group's demand is uncorrelated with the demand of the rest of the aggregation.}

\begin{figure}
	\centering
	\includegraphics[width=.6\columnwidth]{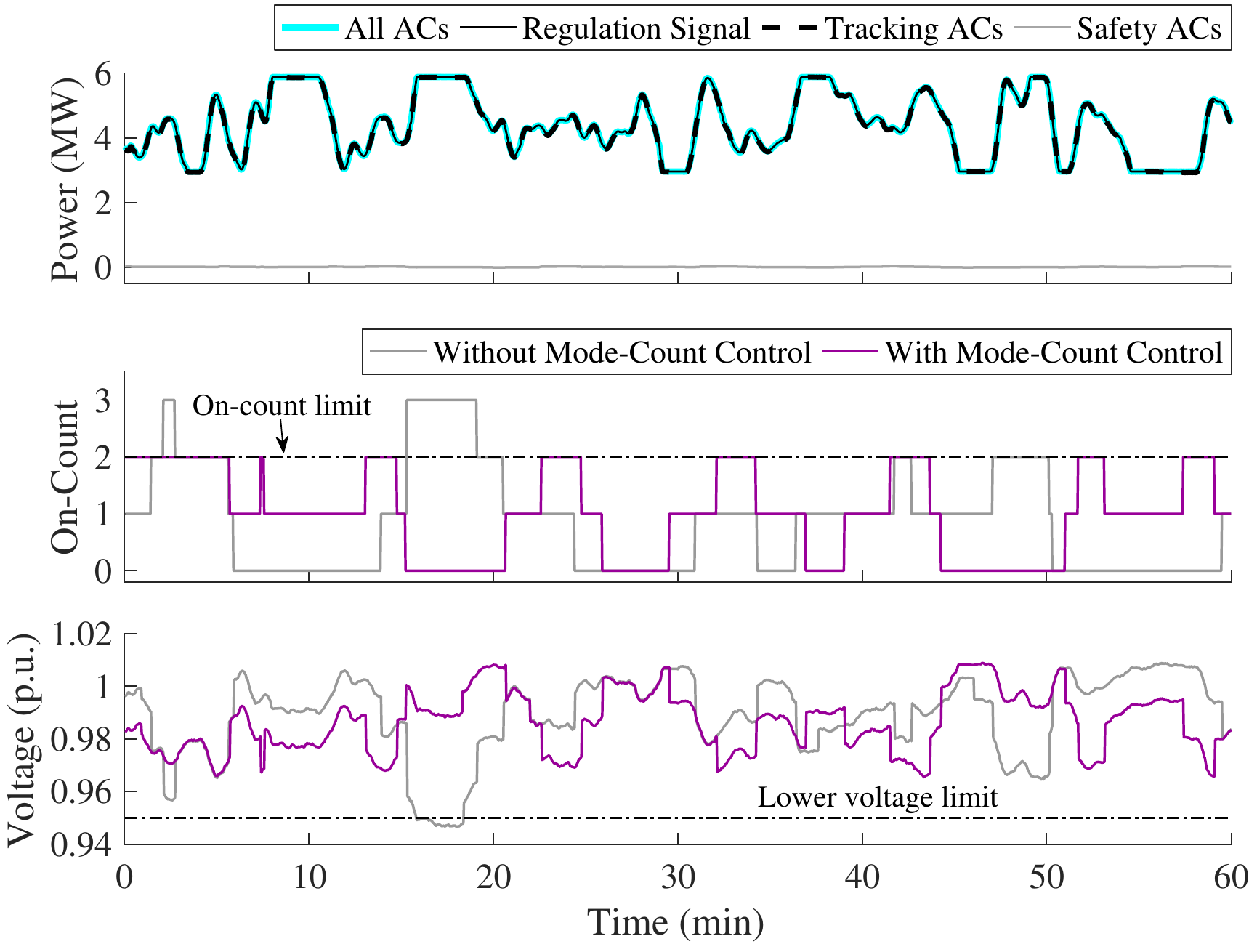}
	\caption{Strategy II during a selected trial. Plot (a) shows the strategy's very high tracking accuracy. The plots in (b) shows that mode-count control can maintain a group of 3 TCLs' on-count to $\leq 2$ (top plot) and thereby prevent under-voltages (bottom plot), here for residential meter \# 469.}
	\label{fig:3plot_stack}
\end{figure}

In Fig.~\ref{fig:3plot_stack}, we demonstrate a few of Strategy~II's key features using the same selected trial. The top plot shows the high-accuracy tracking of Strategy II's individual-model based controller. Compared to Strategy I, Strategy II has very few ACs under safety control---only 0.4\% in this trial. Strategy II also results in fewer switching actions than Strategy I, and therefore, fewer locked ACs. Both of these features make Strategy II less prone to saturation, as can be seen by its accurate tracking in the 6$^\text{th}$ minute. The middle and lower plots of Fig.~\ref{fig:3plot_stack} demonstrate Strategy II's mode-count controller. Without mode-count control, an under-voltage occurs at meter 469 between minutes 15-18. With mode-count control, the three AC's connected to meter 469 are constrained to have an on-count $\leq 2$, and this prevents the under-voltage from occurring.

\begin{figure}
	\centering
	\includegraphics[width=.6\columnwidth]{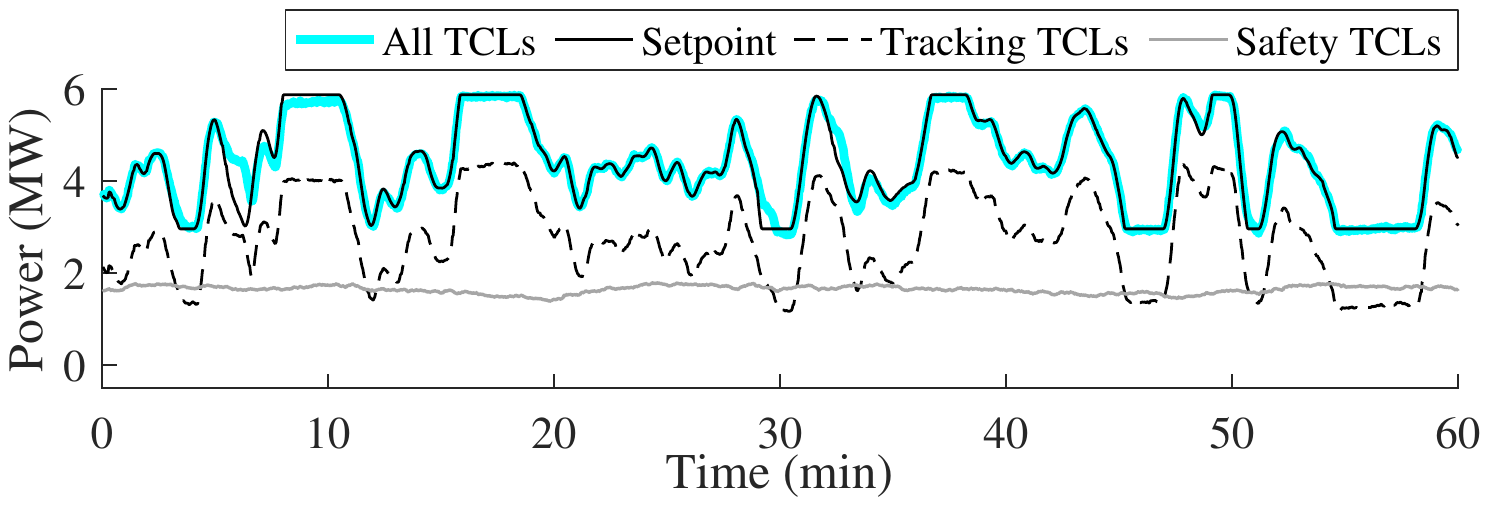}
	\caption{Benchmark strategy during a selected trial. At minutes 6, 8, 29, and 32, there is noticeable tracking error, more so than in Strategies I and II.}
	\label{fig:1plot_PI}
\end{figure}

In Fig.~\ref{fig:1plot_PI}, we show the benchmark strategy's tracking accuracy for the same trial as Figs.~\ref{fig:3plot_ROM} and \ref{fig:3plot_stack}. The plot shows noticeable tracking error in minutes 6, 8, 29, and 32, much more than what occurs in Strategy I or Strategy II. Similar to Strategy I, a large percentage of ACs must be blocked in this trial to ensure network constraints are satisfied.

Overall, we find that tracking performance improves as a control strategy uses more information. In this study, additional information comes in several forms: estimation, modeling, and communication. Strategy I uses state estimation and modeling and outperforms the benchmark strategy.  Strategy~II uses more detailed models and more communication than the other strategies, and has the best tracking performance. Finally, all three strategies perform better when there is direct communication between the operator and aggregator. However, information and direct communication come at a cost. Furthermore, Strategy II requires more computation than Strategy I, which requires more computation than the benchmark strategy, and computation also comes at a cost. In practice, aggregators and operators will need to analyze these trade-offs to determine which strategy best suits their performance/cost requirements.

 %Overall, our results suggest that when a control strategy has more access to information it has better tracking performance. Notably, this increased information can come in several forms: estimation, coordination between aggregator and operator, communication with TCLs, or detailed modeling. Strategy I uses state estimation and modeling and thereby has better tracking performance than the benchmark strategy. All three strategies perform better when there is coordination between the operator and aggregator. Finally, Strategy II utilizes more detailed models and more communication than the other two strategies, and has the best tracking performance. 

\section{Conclusions}

We have proposed two strategies for network-safe load control; in the strategies, a third-party aggregator controls TCLs to provide frequency regulation, while the operator overrides the aggregator's control when necessary to ensure network safety. The two control strategies differ markedly in terms of performance and ease of implementation. Strategy II substantially outperforms Strategy I. Its safety controller is able to prevent network violations using less than 3\% of the ACs used by Strategy~I. This reduction in the number of safety ACs, in combination with a more information-rich tracking controller, results in improved tracking performance for Strategy II. However, Strategy II would likely be more difficult to implement. Compared with Strategy I, it is less scalable in terms of modeling and communication requirements. Choosing between these alternative control strategies will require balancing the costs of implementing the more intensive strategy against the benefits of its improved performance.

%\rnew{There are multiple avenues for future work. First, testing other combinations of controllers (e.g., blocking paired with individual-model based tracking control) would provide additional insight into the controllers' relative merits. Second, the proposed strategies raise questions about how to fairly compensate TCLs that are frequently blocked as a result of being in constrained areas of the network. For example, if TCLs are compensated only for their contributions to frequency regulation, this will disadvantage TCLs that are frequently blocked. Developing fair policies for compensating loads is important future work. Third, we plan to develop and compare additional control strategies with increasing amounts of coordination between aggregator and operator, up to and including a fully centralized solution. Such a comparison will enable operators and aggregators to choose the level of coordination that best suits their privacy, performance, and safety requirements.}

\appendices

\section{Method for selecting TCLs to block or control for safety} \label{app:block}
We first determine which constraints are at risk by running a
preliminary simulation of the network with the TCL aggregation
controlled to provide frequency regulation. Constraints
that experience at least one violation during the simulation
are considered at risk. Next, we select which TCLs to block
to relieve these at-risk constraints; this selection process is
described in detail in the next paragraph. Then we re-run the
simulation with the selected TCLs blocked and check if all
at-risk constraints have been relieved and if any additional
violations have occurred. If there continue to be violations,
additional TCLs are added to the set of blocked TCLs and
the simulation is re-run. This process is repeated until no
violations occur. We note that this offline method assumes advanced knowledge of the regulation signal and a perfect
model of load behavior for the next hour. Thus, the method's
selections (of which TCLs to block) should be interpreted
as best-case selections. An online method for selecting TCLs
would likely result in more TCLs being blocked and may not
successfully avoid all constraint violations.

The method that we use for selecting TCLs depends on
the type of constraint that is at risk. If a line or transformer
are overloaded, then we incrementally block TCLs that are
downstream from the component according to how much they
contribute to the overload. We assume the network is radial
and therefore a TCL's contribution can be approximated as the
sum of its demand and associated line losses. We assume a
TCL’s associated line losses are proportional to the length of
line between the TCL and the overloaded component. Thus,
we block TCLs that are farthest downstream of the component
first. Note that differences in line resistances and differences
in TCL power ratings are neglected. If a service node has a
voltage violation, then we incrementally block TCLs according
to how much they contribute to the voltage drop between
the constrained node and the substation. For loads located
along the line segment between the constrained node and the
substation, the loads' contributions to the voltage drop can be
approximated by their distance along the line segment, with
loads located farther from the substation contributing more
(see Ch. 3, Section 4.1 of \cite{kersting_distribution_2012}). Thus, TCLs located at the
constrained node, or downstream of the node, contribute the
most to the voltage drop and are blocked first. If the violation
persists, then we incrementally block TCLs along the line
segment starting at the constrained node and moving toward
the substation.

\section{Control Policy Derivation} \label{ap:derivation}

We derive the policy by solving the aggregate model \eqref{eq:ltv_kf} for $\bm{u}_t$ given state estimate $\hat{\bm{x}}_t$. To achieve the desired output $\bm{y}^*_{t+1}$, we choose $\bm{u}_t$ such that 
$\bm{y}^*_\text{t+1} = \bm{C} \bm{A}_{\text{u},t}\bm{A}_\text{s}\hat{\bm{x}}_t$.
The matrix $\bm{A}_\text{u}$ can be decomposed into two matrices: $\bm{A}_{\text{u},t} = \tilde{\bm{A}}_{\text{u},t} + \bm{I}$, and therefore
$\bm{y}^*_\text{t+1} = \bm{C} \tilde{\bm{A}}_{\text{u},t}\bm{A}_\text{s}\hat{\bm{x}}_t + \bm{C}\bm{A}_\text{s}\hat{\bm{x}}_t$.
Given the structure of $\bm{A}_{\text{u},t}$, the second row of this equation is satisfied by any choice of $\bm{u}_t$. The first row of this equation is
\begin{equation}
 P^*_{\text{total},{t+1}}\hspace{-3pt} = \overline{p}_\text{on}N \hspace{-2pt}\left(\sum_{n=1}^{N_\text{I}} u_t^n x_{\text{pred,}t}^{n} +\hspace{-6pt}\sum_{n=N_\text{I}+1}^{2N_\text{I}} \hspace{-6pt} -u_t^n x_{\text{pred,}t}^n\right)\hspace{-1pt}  +\bm{C}^1\bm{A}_\text{s}\hat{\bm{x}}_t ,
\label{eq:input2}
\end{equation}
where $P^*_{\text{total},{t+1}}$ is the first entry of $\bm{y}^*_\text{t+1}$ and $\bm{x}_{\text{pred},t}=\bm{A}_\text{s}\hat{\bm{x}}_t$.

%$P^*_{\text{total},{t+1}}$ is the desired aggregate power at time $t+1$
%The right side of \eqref{eq:input2} represents the change in total power that is needed to achieve the desired power in the next time step and relies on a prediction of the total power in the next time step, absent control (i.e., the right-most term).

We design the control policy such that, in each time step, TCLs are switched in only one direction (\emph{on} or \emph{off}). The entries of $\bm{u}_t$ are probabilities and must be between 1 and 0. Given these restrictions, when a positive change in power is needed, \eqref{eq:input2} can be simplified and rearranged such that
\begin{equation}
\sum_{n=1}^{N_\text{I}} u_t^n x_{\text{pred,}t}^{n}  = K \frac{P^*_{\text{total},{t+1}}-\bm{C}^1\bm{A}_\text{s}\hat{\bm{x}}_t}{\overline{p}_\text{on}N},
\label{eq:control_on}
\end{equation}
 where $K$ has been added as a proportional gain. Similarly, when a negative change in power is needed, the last $N_\text{I}$ entries of $\bm{u}_t$ must satisfy
\begin{equation}
\sum_{n=N_\text{I}+1}^{2N_\text{I}} -u_t^n x_{\text{pred,}t}^n  = K \frac{P^*_{\text{total},{t+1}}-\bm{C}^1\bm{A}_\text{s}\hat{\bm{x}}_t}{\overline{p}_\text{on}N},
\label{eq:control_off}
\end{equation}
and all other entries of $\bm{u}_t$ are set to zero.

\section{Network Modifications}\label{ap:feedermod}
We make the following five modifications to improve the accuracy of the network model: 1) we increase a distribution transformer's rating if its average load over an hour is higher than both its original planning load and its original rating; 2) we increase the size of triplex lines if their maximum current is larger than their rating; 3) we shift the regulation range of capacitor banks so that voltages stay within 0.95-1.05 p.u. during nominal operation; 4) we adjust the capacitor bank settings such that phases are individually controlled; 5) we set the initial conditions of capacitors and regulators so that they are at steady state condition when the test trials begin.

%\section*{Acknowledgment}

%The authors thank 

\ifCLASSOPTIONcaptionsoff
  \newpage
\fi



%\bibliographystyle{IEEEtran}
\bibliographystyle{IEEEtran}

\bibliography{sjc_references}


%\newpage


%\vfill


\end{document}